\documentclass[twocolumn,superscriptaddress, aps, prl, showpacs,preprintnumbers,amsmath,amssymb]{revtex4-1}

\usepackage{graphicx}
\usepackage{dcolumn}
\usepackage{bm}
\usepackage{color}
\usepackage{ulem}

\begin{document}

\title{
Anomalous Hall effect from frustration-tuned scalar chirality distribution in Pr$_2$Ir$_2$O$_7$
}

\author{M. Udagawa}%
\affiliation{%
Department of Applied Physics, University of Tokyo, Tokyo 113-8656, Japan
}
\affiliation{%
Max-Planck-Institut f\"{u}r Physik komplexer Systeme, 01187 Dresden, Germany
}%

\author{R. Moessner}
\affiliation{%
Max-Planck-Institut f\"{u}r Physik komplexer Systeme, 01187 Dresden, Germany
}%

\date{\today}

\begin{abstract}
We analyse the Ising Kondo lattice model on a pyrochlore structure in order to 
study the anomalous Hall effect due to non-coplanar magnetism. We focus on the frustration-induced spatial inhomogeneity of different  magnetic low-temperature regimes, between which one can efficiently tune  using an external magnetic field. 
We incorporate non-magnetic scattering on a phenomenological level so that we can distinguish between the effects of short-range correlations and short-range coherence. We obtain a Hall conductivity ($\sigma_{\rm H}$) as function of field strength and direction which compares well to the experimental data of Pr$_2$Ir$_2$O$_7$. In particular, we show that  the observed peak in $\sigma_{\rm H}$ for ${\mathbf H}\parallel[111]$ signals the crossover from zero-field spin ice to Kagome ice.
\end{abstract}

\maketitle
The properties of itinerant degrees of freedom on geometrically frustrated lattices are only poorly understood. One promising avenue for studying the interplay of frustration and itinerance are hybrid systems where itinerant electrons interact with localized magnetic moments subject to strong frustration. The latter can exhibit various exotic phases incorporating peculiar spatial correlations \cite{MoessnerRamirezPT}. It is therefore natural to ask whether these bequeath their unusual behavior to the itinerant electrons, resulting in novel types of behavior for the composite system.
 
The anomalous Hall effect (AHE) is one particularly striking resulting phenomenon\ \cite{Nagaosa10}. AHE was originally associated with ferromagnetic conductors with strong spin-orbit interaction\ \cite{KarplusLuttinger54, Smit55, Berger70}. However, AHE has recently been reinterpreted in a broader context, including non-coplanar magnets as promising candidates for its emergence\ \cite{Ohgushi00,Taillefumier06, Taguchi01,Machida07}.

Prominently, the compound Pr$_2$Ir$_2$O$_7$ shows a unique Hall response. It is composed of two interpenetrating pyrochlore lattices. Ir $5d$ electrons form a conduction band on one, while the localized Pr $4f$ moments reside on the other, and develop spin-ice-type correlation at low temperature\cite{Nakatsuji06, Udagawa12}. It is quite plausible that the spin scalar chirality of the spin ice manifold gives rise to nontrivial features in the Hall response, particularly strikingly in zero field \cite{Machida10}. In addition, the observed Hall conductivity is highly anisotropic and non-monotonic, with a prominent peak around $H\sim\ 0.7$ Tesla for ${\mathbf H}\parallel[111]$\ \cite{Machida07,Balicas11,Machida10}. 
Pioneering analyses of the pyrochlore conductors\ \cite{Kalitsov09,Tomizawa10} have considered spatially periodic structures for the localized moments. It is now natural to ask how spatial aperiodicity -- arising from the geometrical frustration of the spin-ice local moments -- manifests itself in the nontrivial Hall response observed in this compound.

We analyse the dependence of $\sigma_{xy}$ on the coupling between itinerant and local degrees of freedom; field direction and strength; as well as damping which we include for the itinerant electrons to phenomenologically take into account non-magnetic scatterers. Our results compare well with the experimental data for Pr$_2$Ir$_2$O$_7$ with ${\mathbf H}\parallel[100]$ and $[111]$. In particular, we find that the prominent peak observed for ${\mathbf H}\parallel[111]$ can be attributed to the crossover from the zero-field spin ice state to Kagome-ice state: the latter is a state with perfectly field-aligned spins on the triangular layer with the other spins disordered but subject to the ice rule constraint\cite{Hiroi03}, see Supplementary I.

In the remainder of this work, we first present the model and analysis method for the relatively simple Kagome ice model. This is then extended to include the features necessary to make detailed contact with the Pr$_2$Ir$_2$O$_7$ data. 

\begin{figure}[t]
\begin{center}
\includegraphics[width=0.49\textwidth]{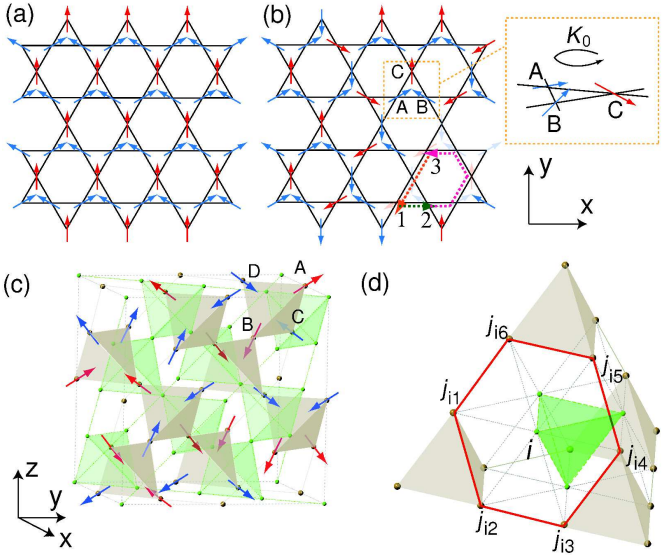}
\end{center}
\caption{\label{latticestructure} 
(color online). (a) A uniform configuration and (b) a representative of the disordered configurations satisfying the Kagome ice rule. Blue (red) arrows show spins ${\mathbf S}_i$ corresponding to $\eta_i=1 (-1)$. An example of the graph belonging to $G[7]$ (see the main text) is shown in (b) with combined three dashed arrows. (c) Structure of double pyrochlore lattice. Green (Gold) tetrahedra constitute the Ir (Pr) pyrochlore lattice. Sublattice indices A, B, C and D are shown for the Pr lattice, for which an example of a spin ice configuration is shown. (d) One Ir tetrahedron surrounded by 4 Pr tetrahedra. Each Ir ion ($i$) has 6 neighboring Pr ions ($j_{i1}\cdots j_{i6}$) forming a hexagon, as highlighted by a thick line.
}
\end{figure}

The Kagome ice model describes localised non-dynamical Ising moments on the Kagome lattice [Fig. \ref{latticestructure} (a), (b)] interacting with itinerant electrons through local fields ${\mathbf h}_i$ at each site:
The Hamiltonian is given by 
\begin{eqnarray}
\mathcal{H} = -t\sum\limits_{\langle i,i'\rangle, \alpha} (c_{i\alpha}^{\dag}c_{i'\alpha} + {\rm H. c.}) - \sum\limits_{i,\alpha,\beta} c_{i\alpha}^{\dag}{\bm\sigma}_{\alpha\beta}c_{i\beta}\cdot{\mathbf h}_i.
\label{Hamiltonian}
\end{eqnarray}
The sum $\langle i,i'\rangle$ is taken over the nearest-neighbor (n.n.) sites. We simply choose ${\mathbf h}_i=J{\mathbf S}_i$, with exchange coupling $J$. The localized spins $\{{\mathbf S}_i\}$ are subject to local easy-axis anisotropy, i.e. ${\mathbf S}_i=\eta_i{\mathbf D}_i$ $(\eta_i=\pm1)$, with ${\mathbf D}_i=\frac{1}{\sqrt{3}}[1, -1, 1], \frac{1}{\sqrt{3}}[1, 1, -1]$ and $\frac{1}{\sqrt{3}}[-1, 1, 1]$, if $i$ belongs to sublattice A, B and C. 
Here, we take a quenched average in terms of $\{{\mathbf S}_i\}$ by imposing ``Kagome ice rule", namely we impose for each triangle $\sum_{i\in\bigtriangleup}\eta_i=1$.

This allows a macroscopic number of spin configurations\ \cite{Udagawa02,Moessner03}, including a uniform configuration [Fig. \ref{latticestructure} (a)], and a huge number of disordered configurations [Fig. \ref{latticestructure} (b)]. Crucially, for each and all of these, the  spin scalar chirality is uniform ${\mathbf S}_a\cdot({\mathbf S}_b\times{\mathbf S}_c)=K_0\equiv -4/3\sqrt{3}$ for all the upward and downward triangles\ \cite{spinscalarchirality}. We can thus examine the effect of spatial disorder on AHE, while preserving uniform spin scalar chirality.

For the calculation of Hall conductivity $\sigma_{xy}$, we randomly generate a series of spin configurations under the Kagome ice rule, $\{{\mathbf S}^{(p)}_i\}$. For each $\{{\mathbf S}^{(p)}_i\}$, the Hall conductivity is given as
\begin{align}
\sigma_{xy}(\{{\mathbf S}^{(p)}_i\})=\frac{e^2}{\hbar V}\sum\limits_{m,m'}&(f(E_{m}) - f(E_{m'}))\nonumber\\
&\times\frac{{\rm Im}(\langle m|J_x|m'\rangle\langle m'|J_y|m\rangle)}{(E_{m} - E_{m'})^2 + 1/\tau^2},
\label{HallConductivity}
\end{align}
by Kubo formula. Here, $|m\rangle$ and $E_m$ are the eigenenergy and corresponding eigenstate of Hamiltonian eq. (\ref{Hamiltonian}). f(E) is the Fermi distribution function at zero temperature. $J_{x(y)}$ is the $x(y)$ component of the current operator, and $V$ is the total volume of the system. Here, we introduce the phenomenological damping rate $1/\tau$ to take account of the finite lifetime of electrons due to non-magnetic impurities. While the magnetic disorder itself causes damping, non-magnetic scattering plays another important role in Hall conductivity. $1/\tau$ sets a coherence length of electrons, which determines the effective spatial scale of spin scalar chirality.  The Hall conductivity $\sigma_{xy}$ can be obtained after taking the configurational average, as $\sigma_{xy}=\frac{1}{N_{\rm s}}\sum_{p=1}^{N_s}\sigma_{xy}(\{{\mathbf S}^{(p)}_i\})$. We typically choose $N_{\rm s}=100$, and system size $N=32\times 32\times 3=3072$ sites. Hereafter, we set $t=\hbar=e^2/h=1$.

\begin{figure}
\begin{center}
\includegraphics[width=0.51\textwidth]{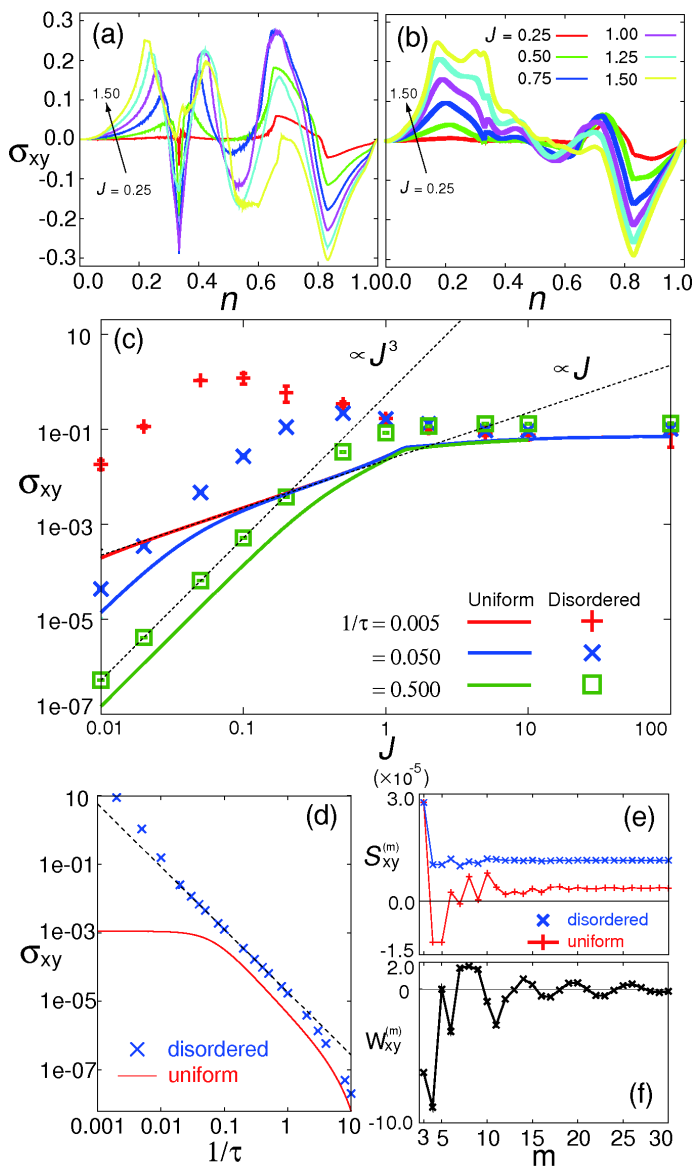}
\end{center}
\caption{\label{kagomeicemodel}
(color online).  Dependence of $\sigma_{xy}$ on particle density, $n$, for (a) the uniform configuration and (b) the disordered configuration at $1/\tau=1.00$. (c) $J$ dependence of $\sigma_{xy}$ at $n=0.0977$ for the disordered case (dots) and the uniform case (solid lines). (d) The $1/\tau$ dependence of $\sigma_{xy}$ at $n=0.0977$ and $J=0.05$. The dashed line is a guide to eye. (e) Partial summation of Hall conductivity $S^{(m)}_{xy}$. (f) The weighting factor $W^{(m)}_{xy}$ averaged over the graphs at each $m$.}
\end{figure}

In Fig.\ \ref{kagomeicemodel}(a) and (b), we show the dependence of $\sigma_{xy}$ on particle density, $n$, in the uniform [Fig. \ref{latticestructure}(a)] and disordered configurations [Fig. \ref{latticestructure}(b)], at relatively large damping, $1/\tau=1.0$. In the uniform case, $\sigma_{xy}$ can be obtained as a summation of Berry curvature in momentum space, showing a steep change around $n\sim 1/3$, where the Dirac cone in the non-interacting band gives a singularity in Berry curvature. This singular behavior is absent in $\sigma_{xy}$ in the disordered case.

$J$ and $1/\tau$ dependence of $\sigma_{xy}$ is summarized in Fig. \ref{kagomeicemodel}(c) at $n=0.0977$. Firstly, for small $J$, a cubic law, $\sigma_{xy}\propto J^3$, is found in both uniform and disordered cases. This cubic law can perturbatively be ascribed to the multiple scattering from triplets of spins exhibiting finite scalar chirality\ \cite{Tatara02}.  With increasing $J$, deviation from the cubic law is found at $J\sim1/\tau$. In particular, in the uniform case, $\sigma_{xy}$ becomes insensitive to $1/\tau$, and another scaling law, $\sigma_{xy}\propto J$, appears, suggesting the $\sigma_{xy}$ is described in terms of the Berry curvature in this region\ \cite{MOnoda04}. For $J\gg t$, the system falls into a double-exchange regime: the itinerant electron spins are aligned with the localized spins, and $\sigma_{xy}$ takes on values only weakly dependent on $1/\tau$.

In general, $\sigma_{xy}$ takes considerably different values between the disordered case ($\sigma^d_{xy}$) and the uniform ordered one, $\sigma^u_{xy}$, as shown in Fig. \ref{kagomeicemodel}(d), where
we plot the $1/\tau$ dependence of $\sigma_{xy}$ at $J=0.05$. This is most pronounced for small damping, $1/\tau\ll J$, where $\sigma^u_{xy}$ saturates, but the difference persists all the way to $1/\tau\gg J$. 

A perturbative treatment in ${\bf h}$ sheds light on the origin of difference between the two cases.
To third order\ \cite{perturbation},
\begin{eqnarray}
\sigma_{xy} = \sum\limits_{(i_1,i_2,i_3)}{\mathbf h}_{i_1}\cdot({\mathbf h}_{i_2}\times{\mathbf h}_{i_3})W_{xy}(i_1, i_2, i_3),
\label{3rdHall}
\end{eqnarray}
where the summation is taken over the $N(N-1)(N-2)/6$ triplets of sites $(i_1, i_2, i_3)$, see Supplementary material II. This gives the Hall conductivity as summation over  the triplets' spin scalar chirality with weighting factor $W_{xy}(i_1, i_2, i_3)$.
It is instructive to resolve the Hall conductivity (\ref{3rdHall}) in the form of a graphical series expansion as $\sigma_{xy} = \sum\limits_{m=3}^{\infty}\sigma^{(m)}_{xy}$.
Here, $\sigma^{(m)}_{xy}$ is the total contribution from the triplets $(i_1, i_2, i_3)$ belonging to the set of graphs $G[m]$ composed of three segments of total length $m$ (An example of a triplet $\in G[7]$ is shown in Fig.\ \ref{latticestructure}(b)). 
 
Fig.\ \ref{kagomeicemodel}(e) and (f) show the partial summation $S^{(m)}_{xy}\equiv\sum_{l=3}^{m}\sigma^{(l)}_{xy}$ and the averaged weighting factor at each $m$, $W^{(m)}_{xy}$, at $J=0.1$ and 
$1/\tau=0.5$. In the disordered case, different contributions for graphs of a given $m\geq5$ come with an effectively random sign, 
hence canceling against one another. By contrast, for the uniform case, the summation continues to oscillate until damping destroys  
coherence at larger $m$. This illustrates the different roles played by loss of correlations of the local moments and loss of coherence of the itinerant electrons in the two respective cases.

Now, let us apply this 3rd-order perturbative scheme to Pr$_2$Ir$_2$O$_7$. Since the exchange coupling $J$ stems from the superexchange process between Pr and Ir ions, it is reasonable to assume $J/t\ll 1$. We consider a double-pyrochlore lattice: two interpenetrating pyrochlore lattices, as shown in Fig. \ref{latticestructure} (c), with itinerant electrons $(c_{i\alpha})$ on the Ir sublattice, and localized Pr moments ($\{\mathbf S[j]\}$) on the other. The localized moments are subject to Ising anisotropy as ${\mathbf S[j]}\equiv\eta_j{\mathbf D}_j$, with ${\mathbf D}_j=\frac{1}{\sqrt{3}}[1,1,1], \frac{1}{\sqrt{3}}[1,-1,-1], \frac{1}{\sqrt{3}}[-1,1,-1]$ and $\frac{1}{\sqrt{3}}[-1,-1,1]$, if $j$ belongs to sublattice A, B, C and D, respectively. Each site $i$ on the Ir sublattice has 6 neighbors ($j_{i1}, j_{i2}\cdots, j_{i6}$) located on the honeycomb ring of the Pr sublattice, as shown in Fig. \ref{latticestructure} (d). To describe the interaction, we adopt the same Ising-Kondo-lattice-type Hamiltonian eq. (\ref{Hamiltonian}) with the local field the sum of the 6 neighboring localized moments ${\mathbf h}_i=J\sum_{l=1}^6{\mathbf S[j_{il}]}$.

While the spin configuration can be determined from the equilibrium condition of Hamiltonian (\ref{Hamiltonian}), we start from the phenomenology that the spin part is described by the nearest-neighbor spin ice Hamiltonian (\ref{SpinHamiltonian}), in order to examine how spin ice correlations in Pr$_2$Ir$_2$O$_7$ affect its Hall conductivity.
\begin{eqnarray}
\mathcal{H}_{\rm spin} = J_{\rm spin}\sum_{\langle j,j'\rangle}\eta_j\eta_{j'} - {\mathbf H}\cdot\sum_j {\mathbf S}[j].\hspace{0.2cm} (J_{\rm spin}>0)
\label{SpinHamiltonian}
\end{eqnarray}
We use $\mathcal{H}_{\rm spin}$ in a standard equilibrium Monte Carlo sampling to obtain $N_s=100$ sets of $\{{\mathbf S}[j]\}$. Though we are interested in the region $T \to 0$, we introduce temperature $T$ as a phenomenological parameter to mimic the deviation from ideal spin ice due to the long-range RKKY interaction in the actual compound, and set $T/J_{\rm spin}=0.5$.
Hereafter, we focus on the field directions ${\mathbf H}\parallel[100]$ and $[111]$\ \cite{Gingras09}. We set magnetic coordinates ${\mathbf e}_x\parallel[010]$ and  ${\mathbf e}_y\parallel[001]$ for  ${\mathbf H}\parallel[100]$, and ${\mathbf e}_x\parallel[\bar{1}10]$ and  ${\mathbf e}_y\parallel[\bar{1}\bar{1}2]$ for ${\mathbf H}\parallel[111]$, and calculate $\sigma_{\rm H}\equiv\sigma_{xy}$. We consider the low density region\ \cite{Machida07}, and fix the particle density at $n=0.01$. As system size, we adopt $N=12\times12\times12\times4=6912$ sites.

\begin{figure}
\begin{center}
\includegraphics[width=0.49\textwidth]{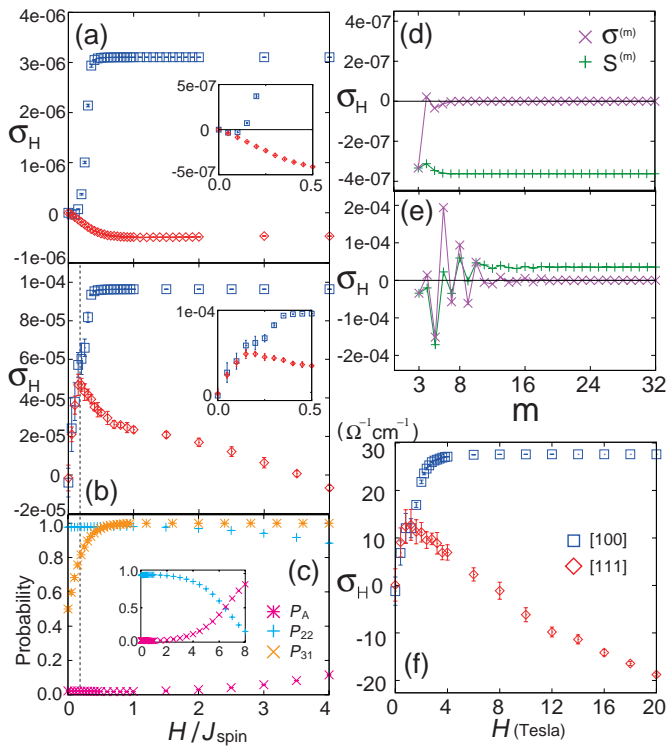}
\end{center}
\caption{\label{doublepyrochloremodel} 
(color online). The magnetic field ($H$) dependence of $\sigma_{\rm H}$ at $n=0.01$ and $J=0.1$ for ${\mathbf H}\parallel[100]$ and $[111]$ at (a) $1/\tau=5.0$ and (b) $0.5$. The results for ${\mathbf H}\parallel[100]$ ($[111]$) are plotted with open squares (diamonds).
The insets are enlarged plots around $H=0$. (c) $H$ dependence of the probabilities $P_{22}$, $P_{31}$, and $P_A$. The crossing of $P_{22}$ and $P_{31}$ is shown in the inset. The vertical dashed line in (b) and (c) is a guide to eye.
(d) (e) Graph-resolved Hall conductivity $\sigma^{(m)}_{\rm H}$ and its partial summation $S^{(m)}_{\rm H}$ at $H/J_{\rm spin}=0.4$ and ${\mathbf H}\parallel[111]$ for (d) $1/\tau=5.0$ and (e) $0.5$. (f) $\sigma_{\rm H}$ of the multi-orbital model. }
\end{figure}

In Figs.\ \ref{doublepyrochloremodel} (a) and (b), we plot the magnetic field dependence of $\sigma_{\rm H}$ at $1/\tau=5.0$ and $0.5$. 
For extremely large damping, $1/\tau=5.0$, only the smallest triangles contribute to $\sigma_{\rm H}$ [Fig.\ \ref{doublepyrochloremodel}(d)]. In this region, the sign of $\sigma_{\rm H}$ becomes opposite between ${\mathbf H}\parallel[100]$ and ${\mathbf H}\parallel[111]$, as expected in Ref.\ \cite{Machida07} on the assumption of the local limit. 
However, the full magnetic field dependence of $\sigma_{\rm H}$, especially the low-field negative linear response in this local limit considerably deviates from the experimental results\ \cite{Machida07} [Fig.\ \ref{doublepyrochloremodel}(a), inset]. 
The negative linear response comes from the large negative contribution at $m=3$ [Fig. \ref{doublepyrochloremodel} (d)]: solely short-range spin ice correlations within the four-Pr cluster [Fig. \ref{latticestructure} (d)]
do not give correct $\sigma_{\rm H}$.

In contrast, for intermediate damping, $1/\tau=0.5$, $\sigma_{\rm H}$ shows quite similar behavior to the experimental data\ \cite{Machida07}. 
For small $H$, $\sigma_{\rm H}$ shows positive linear response irrespective of the field direction\ 
[Fig.\ \ref{doublepyrochloremodel}(b), inset]. 
In Fig. \ref{doublepyrochloremodel} (e), we plot the graph-resolved Hall conductivity at $H=0.4$ for ${\mathbf H}\parallel[111]$. This plot shows that the spatially extended scalar chirality beyond the local limit ($m\lesssim10$, or roughly 35\AA) plays a crucial role in the positive linear response. 

The low-field peak for ${\mathbf H}\parallel[111]$ is the most conspicuous feature of the Hall conductivity in Pr$_2$Ir$_2$O$_7$. In previous studies\ \cite{Machida07,Tomizawa10}, this peak is attributed to the spin flip crossover from the low-field spin ice state with dominant 2-in 2-out configuration to the high-field saturated state with 3-in 1-out and 1-in 3-out spin configurations. 
However, our analysis suggests a different picture. In Fig.\ \ref{doublepyrochloremodel} (c), we plot 
the probabilities that each tetrahedron is occupied by 2-in 2-out configuration ($P_{22}$), and 3-in 1-out or 1-in 3-out configurations ($P_{31}$). This plot shows that $P_{22}$ and $P_{31}$ are almost constant at low fields, until the spin flip crossover happens at much higher field $H\sim 6J_{\rm spin}$\ \cite{Hiroi03} [Fig.\ \ref{doublepyrochloremodel}(c), inset].

The peak of $\sigma_{\rm H}$ seems rather related to the crossover from the zero-field spin ice state to Kagome-ice state.
In Fig.\ \ref{doublepyrochloremodel} (c), we plot the probability that a spin on the sublattice A aligns parallel to the field ($P_A$), as an indicator of the Kagome ice state. $P_A$ changes from $0.5$ at $H=0$ to $1.0$ at the Kagome ice state.
The peak of $\sigma_{\rm H}$ corresponds to $P_A\sim 0.75$, i.e. the midpoint of this saturation process, clearly showing that the peak signals the crossover to a Kagome ice state. 
Within the nearest-neighbor spin ice model used here, the crossover occurs at $H\sim T$. 
Accordingly, the peak is located at $H\sim0.5T$, see Supplementary material I.
Indeed, the $[111]$ magnetization takes $M=M_{\rm p}\sim\frac{1}{3}M_{\rm sat}$ experimentally, when $\sigma_{\rm H}$ has a peak, with $M_{\rm sat}\sim1.5\mu_{\rm B}/$Pr, the saturated magnetization\ \cite{Machida07}. This $M_{\rm p}$ coincides with 
the value at the midpoint of saturation process of $M_A$, and smaller than $M_{\rm s}\sim\frac{5}{6}M_{\rm sat}$ expected at the spin flip crossover.

Here, let us turn to the physical origin of the peak.
As the magnetic field is applied, net spin scalar chirality, and hence  $\sigma_{\rm H}$,  is enhanced. On the other hand, 
the evolution to the Kagome ice state can be regarded as a partial ordering process of sublattice A, so that spatial disorder is reduced, suppressing the interference between the graphs resulting in a suppression of
$\sigma_{\rm H}$, as discussed above for Kagome ice model.
It is tempting to note that
the balance between the two effects gives rise to a prominent peak during the evolution to the Kagome ice state.
Indeed, the peak becomes more prominent as $1/\tau$ is further reduced (not shown), reinforcing the subtle balance between the 
two competing effects. 
The sensitivity to $1/\tau$ may be confirmed from the systematic study of sample dependence of $\sigma_{\rm H}$ and resistivity. 
Further analyses are clearly desirable to elucidate this point.

We finally comment on the quantitative aspect of our theory. Although our result is based on a single orbital tight-binding model, the overall features of $\sigma_{\rm H}$ are insensitive to band structure. However, the amplitude of the Hall conductivity is sensitive to `details' of the band structure. By adopting a realistic band structure based on a multi-orbital tight-binding model with three $t_{2g}$ orbitals, we could obtain the Hall conductivity $\sim 30\Omega^{-1}$cm$^{-1}$ at high fields, comparable to experimental data [Fig. \ref{doublepyrochloremodel} (f)], with reproducing various features of experimental results; See supplementary material III.

The authors thank for fruitful discussions with S. Nakatsuji, Y. Machida and Y. Motome.
This work was supported by KAKENHI (Nos. 24740221 \& 23102708 \& 24340076).
The computation in this work has been partially done using the facilities of the
Supercomputer Center, Institute for Solid State Physics, University of Tokyo.

\setcounter{figure}{0}
\begin{widetext}
\vspace*{0.3cm}
\begin{center}
\Large 
Supplementary Items for {\it Anomalous Hall effect from frustration-tuned scalar chirality distribution in Pr$_2$Ir$_2$O$_7$}
\end{center}

\section{I. Crossover between the spin ice and the kagome ice state}
\begin{figure}[h]
\begin{center}
\includegraphics[width=0.49\textwidth]{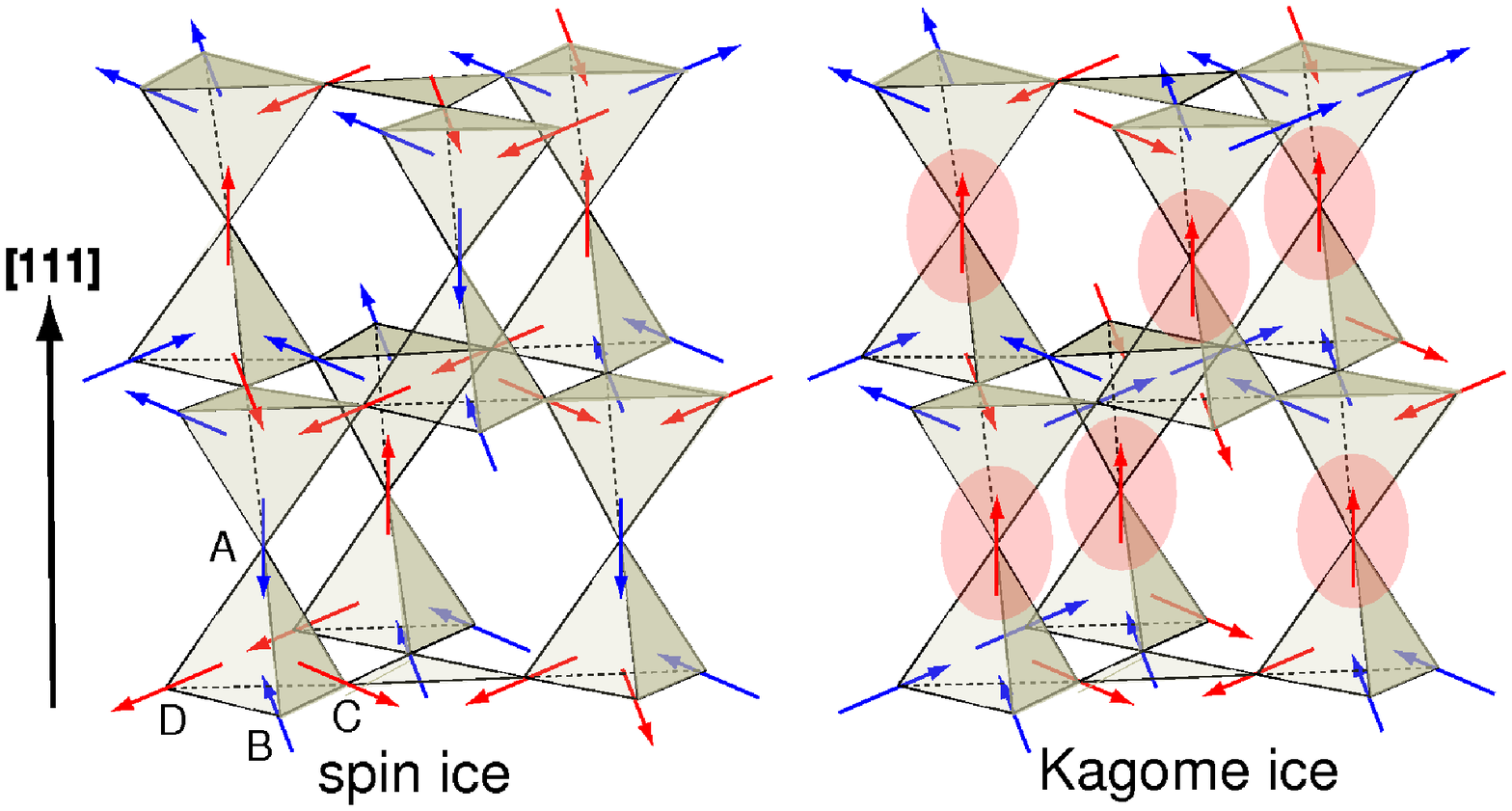}
\end{center}
\caption{\label{Figure_supplementary} 
(color online). A schematic picture of the crossover from zero-field spin ice to Kagome ice. Only the Pr lattice is shown. In Kagome ice the spins are field-aligned on sublattice A indicated by shaded ellipses, which form layered triangular lattices.
 }
\end{figure}

\begin{figure}[h]
\begin{center}
\includegraphics[width=0.49\textwidth]{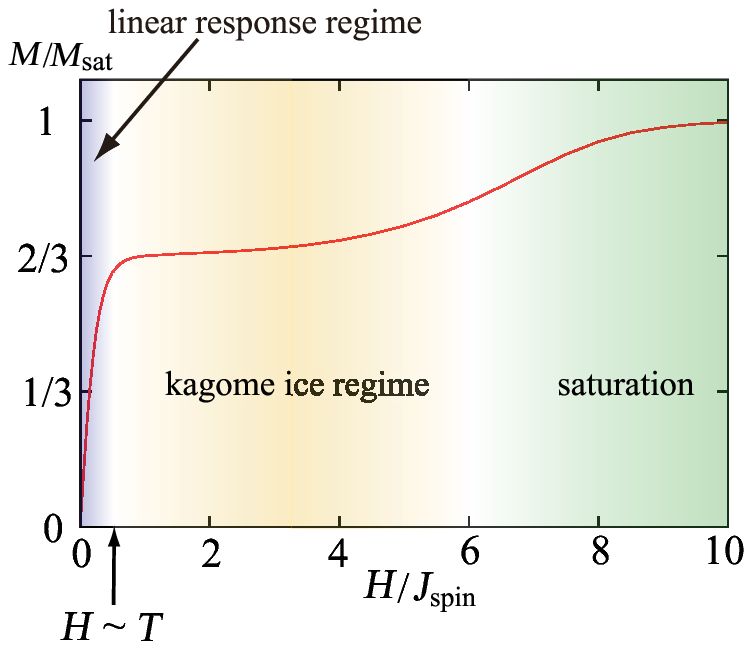}
\end{center}
\caption{\label{Figure_supplementary2} 
(color online). The magnetization process of the nearest-neighbor spin ice model eq. (4) in the main text, obtained at $T=0.5J_{\rm spin}$.
The magnetization ($M$) shows different behaviors at each magnetic field ($H$) regions. 
 }
\end{figure}
The spin ice state is characterized by the so-called ice rule: out of the four spins on each tetrahedra, two point inward while the other two point outward, satisfying the 2-in 2-out constraint [Fig. \ref{Figure_supplementary}]. Under the magnetic field ${\mathbf H}\parallel[111]$, the spin ice state changes to the kagome ice state, which has perfectly field-aligned spins on the triangular layer with the other spins disordered but subject to the ice rule [Fig. \ref{Figure_supplementary}]. Looking from the $[111]$ direction, a pyrochlore lattice can be considered as the alternate stacking of kagome and triangular lattices. Since the spins on the triangular layers have easy axis parallel to $[111]$, these spins tend to
align prior to those on the kagome lattice under the magnetic field ${\mathbf H}\parallel[111]$.

In Fig. \ref{Figure_supplementary2}, we plot the magnetic field ($H$) dependence of magnetization ($M$) of the nearest-neighbor spin ice model eq. (4) in the main text, obtained at $T=0.5J_{\rm spin}$. For $H\lesssim T$, $M$ grows almost linearly, and reaches $M=\frac{2}{3}M_{\rm sat}$ at $H\sim T$, with $M_{\rm sat}$ the saturated value of magnetization for ${\mathbf H}\parallel[111]$. For $T\lesssim H\lesssim6J_{\rm spin}$, the kagome ice state is stabilized with nearly constant magnetization $M\sim\frac{2}{3}M_{\rm sat}$. For $H\gtrsim6J_{\rm spin}$, the spins on the kagome layers also align in the field direction. As a result, the magnetization gradually approaches its saturated value, $M\sim M_{\rm sat}$.

\section{II. Third-order perturbation theory of anomalous Hall conductivity}
In this section, we give a detailed derivation of the perturbative formula of Hall conductivity [eq.\ (3) in the main text].
We start with the Ising Kondo Hamiltonian [eq. (1) in the main text], in momentum space representation,
\begin{eqnarray}
\mathcal{H} = \sum\limits_{{\mathbf k}, s}\sum\limits_{a,a'}{\mathcal H}_{aa'}({\mathbf k})c^{\dagger}_{{\mathbf k}as}c_{{\mathbf k}a's}
- \frac{J}{N_{\rm c}}\sum\limits_{{\mathbf k}, {\mathbf k}'}\sum\limits_{ss'}\sum\limits_{a}{\mathbf h}_{{\mathbf k}-{\mathbf k}', a}\cdot(c^{\dagger}_{{\mathbf k}as}\bm{\sigma}_{ss'}c_{{\mathbf k}'as'})
\equiv \mathcal{H}_0 + \mathcal{H}'.
\label{Hamiltonian_k}
\end{eqnarray}
Here, we have introduced sublattice indices $a$ and $a'$. $N_{\rm c}$ is the number of unit cells.
${\mathcal H}_{aa'}({\mathbf k})$ is the kinetic energy in momentum representation. For example, for the Kagome ice model, it is given by
\begin{eqnarray}
{\mathcal H}_{aa'}({\mathbf k}) = 
\begin{bmatrix}
0 & -2\cos({\mathbf k}\cdot({\mathbf e}_{\rm A} - {\mathbf e}_{\rm B})) &  -2\cos({\mathbf k}\cdot({\mathbf e}_{\rm A} - {\mathbf e}_{\rm C}))\\
-2\cos({\mathbf k}\cdot({\mathbf e}_{\rm B} - {\mathbf e}_{\rm A})) & 0 &  -2\cos({\mathbf k}\cdot({\mathbf e}_{\rm B} - {\mathbf e}_{\rm C}))\\
-2\cos({\mathbf k}\cdot({\mathbf e}_{\rm C} - {\mathbf e}_{\rm A})) &  -2\cos({\mathbf k}\cdot({\mathbf e}_{\rm C} - {\mathbf e}_{\rm B})) & 0
\end{bmatrix}.
\label{Hk_kagome}
\end{eqnarray}
Here, ${\mathbf e}_{\rm A}$, ${\mathbf e}_{\rm B}$ and ${\mathbf e}_{\rm C}$ are the internal coordinates in a unit cell, corresponding to sublattice A, B and C.
We consider the current correlation function
\begin{eqnarray}
Q^{\mu\nu}(i\omega_q) = \frac{1}{V}\int\limits_0^{\beta}\ e^{i\omega_q\tau}\langle J_{\mu}(\tau)J_{\nu}(0)\rangle,
\end{eqnarray}
defined with the current operator
\begin{eqnarray}
J_{\mu} = \sum_{{\mathbf k}, a, a', s}\frac{\partial\mathcal{H}_{aa'}({\mathbf k})}{\partial k_{\mu}}c^{\dag}_{{\mathbf k}as}c_{{\mathbf k}a's}.
\end{eqnarray}
and Matsubara frequency $\omega_q = 2\pi qT$ ($q$: integer).
The Hall conductivity $\sigma^{\mu\nu}$ can be obtained as an asymmetric part of  the derivative of correlation function as
\begin{eqnarray}
\sigma^{\mu\nu} = \frac{d}{id\omega}\frac{Q^{\mu\nu}(\omega+i0) - Q^{\nu\mu}(\omega+i0)}{2} = \lim_{T\to0}\frac{1}{4\pi T}(Q^{\nu\mu}(2\pi iT) - Q^{\mu\nu}(2\pi iT)).
\label{defHall}
\end{eqnarray}
We expand $Q^{\mu\nu}(i\omega_q)$ to third order of $\mathcal{H}'$,
\begin{align}
&Q^{\mu\nu}(i\omega_q)\simeq Q^{\mu\nu(3)}(i\omega_q)\nonumber\\
&=\frac{J^3}{6V}\int\limits_0^{1/T}d\tau\ e^{i\omega_q\tau}\int\limits_0^{1/T}d\tau_1\int\limits_0^{1/T}d\tau_2\int\limits_0^{1/T}d\tau_3\sum\limits_{{\mathbf k}sab}\sum\limits_{{\mathbf k}'s'a'b'}\frac{\partial\mathcal{H}_{ab}}{\partial k^{\mu}}\frac{\partial\mathcal{H}_{a'b'}}{\partial k^{\nu}}\sum\limits_{a_1a_2a_3}\sum\limits_{\alpha\beta\gamma}\frac{1}{N}\sum\limits_{{\mathbf k}_1{\mathbf k}_1'}\frac{1}{N}\sum\limits_{{\mathbf k}_2{\mathbf k}_2'}\frac{1}{N}\sum\limits_{{\mathbf k}_3{\mathbf k}_3'}\nonumber\\
&\times h^{\alpha}_{{\mathbf k}_1-{\mathbf k}_1',a_1}h^{\beta}_{{\mathbf k}_2-{\mathbf k}_2',a_2}h^{\gamma}_{{\mathbf k}_3-{\mathbf k}_3',a_3}\sigma_{s_1s_1'}^{\alpha}\sigma_{s_2s_2'}^{\beta}\sigma_{s_3s_3'}^{\gamma}\nonumber\\
&\times\langle\mathcal{T}_{\tau}c^{\dagger}_{{\mathbf k}as}(\tau)c_{{\mathbf k}bs}(\tau)c^{\dagger}_{{\mathbf k}'a's'}(0)c_{{\mathbf k}'b's'}(0)c^{\dagger}_{{\mathbf k}_1a_1s_1}(\tau_1)c_{{\mathbf k}_1'a_1s_1'}(\tau_1)c^{\dagger}_{{\mathbf k}_2a_2s_2}(\tau_2)c_{{\mathbf k}_2'a_2s_2'}(\tau_2)c^{\dagger}_{{\mathbf k}_3a_3s_3}(\tau_3)c_{{\mathbf k}_3'a_3s_3'}(\tau_3)\rangle\nonumber\\
&= -\frac{4}{V}\frac{J^3}{N^3}\sum\limits_{aba'b'}\sum\limits_{a_1a_2a_3}\sum\limits_{{\mathbf k}{\mathbf k}'}\frac{\partial\mathcal{H}_{ab}}{\partial k^{\mu}}\frac{\partial\mathcal{H}_{a'b'}}{\partial k^{\nu}}\sum\limits_{{\mathbf k}''}{\mathbf h}_{{\mathbf k}-{\mathbf k}'',a_1}\cdot({\mathbf h}_{{\mathbf k}''-{\mathbf k}',a_2}\times{\mathbf h}_{{\mathbf k}'-{\mathbf k},a_3})\nonumber\\
&\times {\rm Im}\Bigl[T\sum\limits_{\epsilon_p}G_{{\mathbf k}ba_1}(i\epsilon_p+i\omega_q)G_{{\mathbf k}''a_1a_2}(i\epsilon_p+i\omega_q)G_{{\mathbf k}'a_2a'}(i\epsilon_p+i\omega_q)G_{{\mathbf k}'b'a_3}(i\epsilon_p)G_{{\mathbf k}a_3a}(i\epsilon_p)\Bigr],
\label{derive3rd}
\end{align}
where we have introduced the bare Green's function,
\begin{eqnarray}
G_{{\mathbf k}a_1a_2}(i\epsilon_p) = \sum_{\alpha}\frac{\langle a_1|u_{\alpha}({\mathbf k})\rangle\langle u_{\alpha}({\mathbf k})|a_2\rangle}{i\epsilon_p - (\epsilon_{\alpha}({\mathbf k}) - \mu) + \frac{i}{\tau}{\rm sign}(\epsilon_p)}.
\end{eqnarray}
with $|u_{\alpha}({\mathbf k})\rangle$ and $\epsilon_{\alpha}({\mathbf k})$ the $\alpha$-th eigenfunction and eigenenergy of ${\mathcal H}_{aa'}({\mathbf k})$. $1/\tau$ is a phenomenological damping rate,
and $\epsilon_p=(2p+1)\pi T$ is a fermionic Matsubara frequency with integer $p$.
By combining eq. (\ref{defHall}) and (\ref{derive3rd}) and making Fourier transformation back to real space, we end up with the following expression
\begin{eqnarray}
\sigma_{xy} = \sum\limits_{(i_1,i_2,i_3)}{\mathbf h}_{i_1}\cdot({\mathbf h}_{i_2}\times{\mathbf h}_{i_3})W_{xy}(i_1, i_2, i_3),
\label{3rd_order_formula}
\end{eqnarray}
with
\begin{align}
W_{xy}&(i_1, i_2, i_3)\nonumber\\ 
&= \lim_{T\to0}\frac{4}{V}\sum\limits_{P(i_1,i_2,i_3)}(-1)^P\sum_{\epsilon_p}{\rm Im}\Bigl[I_x(i\epsilon_p, i\omega_q, i_1,i_2)J(i\epsilon_p+i\omega_q, i_2,i_3)I_y(i\epsilon_p+i\omega_q, -i\omega_q, i_3,i_1)\Bigr],
\label{Kernel}
\end{align}
where $\omega_q$ is set to be $\omega_q=2\pi T$.
In eq. (\ref{Kernel}), the summation is taken over the $6$ permutation of sites $i_1, i_2$ and $i_3$ with the sign of permutation $(-1)^P$.
The weighting factor $W_{xy}(i_1, i_2, i_3)$ contains information about the electronic structure of the system through the functions $I_{\mu}\equiv{\mathbf I}\cdot{\mathbf e}_{\mu}$ and $J$: 
\begin{eqnarray}
\left\{\begin{array}{ll}
{\mathbf I}(i\epsilon_p, i\omega_q; i_1,i_2) = \frac{1}{N}\sum\limits_{\mathbf k}\sum\limits_{a,a'}G_{{\mathbf k}a_1a}(i\epsilon_p)\frac{\partial{\mathcal H}_{aa'}({\mathbf k})}{\partial{\mathbf k}}G_{{\mathbf k}a'a_2}(i\epsilon_p+i\omega_q)e^{i{\mathbf k}\cdot({\mathbf r}_{i_1} - {\mathbf r}_{i_2})},\\
J(i\epsilon_p; i_1,i_2) = \frac{1}{N}\sum\limits_{\mathbf k}G_{{\mathbf k}a_1a_2}(i\epsilon_p)e^{i{\mathbf k}\cdot({\mathbf r}_{i_1} - {\mathbf r}_{i_2})},
\end{array}\right.
\end{eqnarray}
where $a_1$ and $a_2$ are the sublattice indices corresponding to sites $i_1$ and $i_2$.

\section{III. Quantitative consistency with experiments}
While our approach based on the single-orbital model gives qualitative description of the Hall conductivity of Pr$_2$Ir$_2$O$_7$,
it requires a realistic band structure to achieve a quantitative agreement with experimental data.
In particular, the amplitude of Hall conductivity is sensitive to orbital degeneracy.
 In Pr$_2$Ir$_2$O$_7$, the conduction bands are composed of three-fold degenerate $t_{2g}$ orbitals of Ir $5d$ electrons. In our perturbative framework, orbital degeneracy can be incorporated by replacing the summation over sites in eq. (3) in the main text, by the summation over both sites and orbitals. This leads to an increase of transverse scattering channels, potentially enhancing $\sigma_{\rm H}$  considerably. 
 
To consider the effect of orbital degeneracy, we adopt a multi-orbital tight-binding model with three $t_{2g}$ orbitals, with Slater Koster determination of transfer integrals via Ir-O-Ir hopping paths. This tight-binding model is equivalent to the one used in ref.\cite{PesinBalents10}. We choose an electron density close to that of $d^5$ configurations in the three-fold degenerate $t_{2g}$ orbitals expected in Pr$_2$Ir$_2$O$_7$: $n=0.85$. As parameters relevant to Pr$_2$Ir$_2$O$_7$, we choose $1/\tau=0.05$, $J=0.073$,
the lattice parameter $10$\AA, $J_{\rm spin}=3.48$K and the magnetic moment of Pr$^{3+}$ multiplet as $3.0\mu_{\rm B}$.
This choice of parameter sets gives a diagonal resistivity $\rho\simeq8.0\times10^2\Omega^{-1}$cm$^{-1}$ within the first Born approximation, of the same order as the experimental value $\sim 2.8\times10^3\Omega^{-1}$cm$^{-1}$\cite{Nakatsuji06,Balicas11}.

As a result, we obtained Hall conductivity as large as $\sim 30\Omega^{-1}$cm$^{-1}$ at high fields ($H\gtrsim 4.0$ Tesla) for ${\mathbf H}\parallel[100]$, comparable to the experimental data, in which the plateau of $\sigma_{\rm H}$ is observed at $H\gtrsim 4.0$ Tesla, taking an almost constant value $\sim 30\Omega^{-1}$cm$^{-1}$. Furthermore, for ${\mathbf H}\parallel[111]$, the Hall conductivity has a peak around $H\sim1.0$ Tesla, and shows sign reversal at $H\sim8.0$ Tesla, again consistent with experimental data where the peak is found at $0.7$ Tesla\cite{Machida07}, while the sign reversal takes place around $H\sim6.0$ Tesla\cite{Balicas11}.

\end{widetext}

\begin{thebibliography}{9}
\bibitem{spinscalarchirality} Here the sites $a$, $b$ and $c$ are chosen in a counterclockwise fashion.
\bibitem{perturbation} $\sigma_{xy}$ in this perturbative scheme does not precisely correspond to the Hall conductivity obtained from the numerical diagonalization method, due to the different treatment of the damping parameter $1/\tau$.
\end{thebibliography}
\end{document}